\documentclass[aps,prb,twocolumn,showpacs,superscriptaddress,floatfix]{revtex4-1}
\pdfoutput=1

\usepackage{amsmath,amssymb}
\usepackage{graphicx}
\usepackage{color}
\usepackage{rotating}
\usepackage{bm}
\usepackage{setspace}
\usepackage{hyperref}
\usepackage{float}
\usepackage{soul}
\usepackage{color}

\hypersetup{
	colorlinks=true,
	urlcolor= blue,
	citecolor=blue,
	linkcolor= blue,
	bookmarks=true,
	bookmarksopen=false,
}	

\begin{document}
	
	\preprint{APS/123-QED}
	
	\title{Twistronics: Manipulating the Electronic Properties of Two-dimensional Layered Structures through their Twist Angle}
	\author{Stephen Carr}
	\affiliation{Department of Physics, Harvard University, Cambridge, Massachusetts 02138, USA.}
	\author{Daniel Massatt}
	\affiliation{School of Mathematics, University of Minnesota, Minneapolis, Minnesota, 55455, USA.}
	\author{Shiang Fang}
	\affiliation{Department of Physics, Harvard University, Cambridge, Massachusetts 02138, USA.}
	\author{Paul Cazeaux}
	\affiliation{School of Mathematics, University of Minnesota, Minneapolis, Minnesota, 55455, USA.}
	\author{Mitchell Luskin}
	\affiliation{School of Mathematics, University of Minnesota, Minneapolis, Minnesota, 55455, USA.}
	\author{Efthimios Kaxiras}
	\affiliation{Department of Physics, Harvard University, Cambridge, Massachusetts 02138, USA.}
	\affiliation{John A. Paulson School of Engineering and Applied Sciences, Harvard University, Cambridge, Massachusetts 02138, USA.}
	\date{\today}
	
	\begin{abstract}
	The ability in experiments to control the relative twist angle between 
	successive layers in two-dimensional (2D) materials
	offers a new approach to manipulating their electronic properties; we refer to this 
	approach as ``twistronics''.  
	A major challenge to theory is that, for arbitrary twist angles, the resulting structure involves 
	incommensurate (aperiodic) 2D lattices. 
	Here, we present a general method for the calculation of the electronic density of states
of aperiodic 2D layered materials, using parameter-free hamiltonians derived from {\it ab initio} 
	density-functional theory. 
	We use graphene, a semimetal, and MoS$_2$, a representative of the transition metal dichalcogenide (TMDC) family of 2D semiconductors, to illustrate the 
	application of our method, which enables fast and efficient simulation of multi-layered stacks 
	in the presence of local disorder and external fields. 
We comment on the interesting features of their Density of States (DoS) as a function of twist-angle and local configuration and on how these features 
	can be experimentally observed. 
		
	\end{abstract}
	
	\maketitle
	
\section{\label{sec:level1}INTRODUCTION}

A few short years after the experimental demonstration 
of the existence of monolayer graphene \cite{Novoselov2004}, 
many other 2D materials,
have been successfully fabricated \cite{Ayari2007, Dean2010, Mak2010, Radisavljevic2011, De2013}. 
Although single-layer 2D systems have intriguing physical properties, 
there has also been great interest in developing and understanding artificial heterostructures
composed of multiple atomic layers 
weakly bonded by van der Waals forces~\cite{Geim2013}.
Mechanical or chemical exfoliation and positioning of 
one layer on top of another allows for a relative twist between successive layers, 
which can destroy the alignment and thereby break the translational symmetry in the
combined system~\cite{Cao2016,Koren2016}.   
The resulting structures may have commensurate stacking for special orientations, but 
more generally are incommensurate.  This allows for interesting new behavior:
studies of bilayer graphene have found clear twist-dependent features 
in both the electronic density of states and the conductivity \cite{Rozhkov2016,Uchida2014}; 
at very small twist-angles, a domain-wall phase appears, related to the 
stacking configuration~\cite{Woods2014}. 
Similar effects may occur in TMDC semiconductors, with 
their band-gaps affected by the substrate 
and the relative twist-angle orientation \cite{Ebnonnasir2014}.  
Incommensurate structures pose a great challenge to 
theoretical studies since the standard description of solids
with crystalline order, a periodic Bravais lattice 
and the associated Bloch states of electrons, is entirely absent in the combined
system although each layer may still be a perfect 2D crystal.  

In the effort to capture the physics of incommensurate systems,
a simple approximation 
is to consider large super-cells that can mimic the incommensurate system;
in the case of first-principles calculations like 
density functional theory (DFT), that can afford relatively small cells, this 
approximation limits the physical system rather severely to special values of the 
twist angle \cite{Uchida2014}. 
This leaves important questions unaddressed:
Are there distinct physical characteristics that distinguish the incommensurate from 
the commensurate case? Do the properties of commensurate systems 
approach the proper limit of the  incommensurate systems as the twist-angle 
is varied? 

In the present work we introduce a robust 
framework for the calculation of the properties of truly 
incommensurate 2D heterostructures that can address such questions
for situations involving arbitrary twists between successive layers.  
Our method is inspired by previous mathematical works on disordered tight-binding models, which can be classified into two distinct concepts.  First, an algebraic treatment of electronic transport in disordered systems~\cite{Bellissard1994,Prodan2012} that allows for
a rigorous definition of quantum-mechanical operators in a disordered material. 
Second, the fact that local tight-binding models create exponentially localized
observables, that is, they make it possible to controllably remove
finite-size and edge effects from calculations~\cite{Chen2015}. We have already provided a rigorous mathematical discussion of this method \cite{Massatt2016}, but here investigate its implications and results for physical systems.
Our modeling is based on effective tight-binding hamiltonians without any 
adjustable parameters, obtained from first-principles DFT results~\cite{Fang2016,Fang2015}. 
As a demonstration of the capabilities of the method, 
we study some prototypical systems of 2D stacked layers, 
including bilayer graphene, a semimetal, and bilayer MoS$_2$, a representative semiconductor 
of the TMDC family. 

\section{\label{sec:level2}FORMALISM}

The essence of our approach consists of the following ideas:
A tight-binding model in $d$-dimensions is described by localized orbitals $\phi_i$ 
in a $d$-dimensional lattice, $i \in \mathbb{Z}^d$, 
and the hopping matrix elements between them labeled $t_{ij}$. 
To describe disorder in this model, we consider the space of all possible defects and calculate physical properties for a carefully chosen subset of configurations. 
This is formulated by defining a configuration space $\Omega$ with specific 
local configurations $\omega \in \Omega$ with a probability distribution $dP(\omega)$. $\Omega$ describes all possible environments that an atom in the infinite crystal can 
experience, and we simulate physical observables 
by sampling over this space of disordered configurations. This is in contrast to periodic approaches, which instead use the Bloch wavenumber, $\vec{k}$, as the sampling space. 
In incommensurate systems translational symmetry has been completely broken, 
and there is no Brillouin zone. 
$\Omega$, referred to as the ``non-commutative Brillioun zone'' for this reason~\cite{Bellissard1994}, is an alternative to this notion; 
neither $\Omega$ nor the Brillioun zone provide a diagonalized band structure with a finite number of eigenvalues at each point.

Viewing the interlayer interaction as a perturbative potential, 
the relative twist-angle can be interpreted as an aperiodic disorder field 
applied to the single-layer system.
For a fixed twist angle, the location of the orbital $\phi_i$ 
in the field created by another layer varies. 
This variation in location can be completely described by the offset, or shift, between the two layers' unit cells, and thus $\Omega$ can be viewed as the compact two-dimensional space of all shifts.
For each shift, we construct a system of finite radius which contributes a finite-size error. 
The error decays exponentially with the radius, so it can be made to approach zero
in a controlable fashion. 
Our results prove that this is a computationally feasible strategy.
	
In this picture, the difference between an incommensurate and commensurate twist angle 
becomes trivial: a commensurate angle has a finite number of possible configurations
because a periodic super-cell exists, 
while an incommensurate angle has an infinite number. 
If two twist angles, $\theta$ commensurate and $\theta'$ incommensurate, 
are extremely close then a specific shift configuration will look effectively identical between them. 
Therefore, the results of a single $\omega$ calculation will not vary significantly between $\theta$ and $\theta'$; rather, it is the sampling of $\omega$ that varies. 
For the twist angle to act as the order parameter in a phase transition
between commensurability and incommensurability, 
a physical observable must vary strongly enough over $\Omega$ 
and the commensurate twist angle must not sample $\Omega$ too finely. 
This distinction only holds for each layer being a perfect infinite crystal. 
In real materials, the difference between an incommensurate and 
a commensurate twist-angle is less clear, as  
the presence of imperfections (strain, tears, ripples) may make even a 
commensurate system sample $\Omega$ continuously. 
%Even at $0^\circ$ twist-angle bilayer materials,
%corresponding to the lowest-energy configuration like the $A-B$ stacking 
%for honeycomb lattices, 
%the observation of domain-wall structures shows that more than a single configuration 
%is present in the real material \cite{Woods2014}. 
%Our method allows us to understand how each configuration contributes to the observed %electronic structure.
		
Our approach can also handle other sources of disorder straightforwardly. 
Magnetic and electric fields can be easily introduced through a Peierl's substitution or 
an on-site energy term, respectively. 
Physical defects such as vacancies, ripples, and edges are easy to implement,
provided that it has been established how the hopping terms 
of the tight-biding hamiltonian change in the presence of defects. 
This is handled by introducing extra dimensionality to $\Omega$ to represent 
all possible forms of disorder and applying them directly in each 
$\omega$ tight-binding model.

Our implementation of these ideas on a high-performance computing system are as follows:\\
i) Creation of a heterostructure model out of layers that are disks of radius $R$;  these 
disks are centered at a point with ``zero-shift'', 
which is just one specific $\omega$ configuration.\\ 
ii) Determination of all relevant hopping indices $H_{ij}$ 
in the sparse hamiltonian by only looking for pairs of orbitals 
that are within the range of the hopping matrix elements $t_{ij}$. \\
iii) For each desired configuration $\omega$, 
displacement of one layer by the some amount 
with respect to the other layer, 
and computation of $H^{\omega}_{ij}$ for each non-zero hopping term; 
from this, we then calculate the local electronic density of states (LEDoS), or any 
other useful physical property like the conductivity.
The LEDoS is derived from the global EDoS, $g(\epsilon)$, 
by considering all eigenstates (indexed by $s$) and orbitals (indexed by $x$):

\begin{equation}
g(\epsilon) = 
%\frac{1}{N} \sum_{s = 1}^{N} \delta(\epsilon - \epsilon_s) = 
%\frac{1}{N} \sum_{s = 1}^{N} \delta(\epsilon - \epsilon_s) \sum_x |\phi_s(x)|^2 = 
\sum_x \frac{1}{N} \sum_{s = 1}^{N} \delta(\epsilon - \epsilon_s)|\phi_s(x)|^2 = \sum_x g_x(\epsilon)
\end{equation}
%where we have identified the local density of states $g_x(\epsilon)$ as:		
%		\begin{equation}
%		g_x(\epsilon) = \frac{1}{N} \sum_{s = 1}^{N} \delta(\epsilon - \epsilon_s)|\phi_s(x)|^2
%		\end{equation}
iv) Application of the operator of interest to $H^{\omega}_{ij}$ 
with a Kernel Polynomial Method (Chebyshev polynomials) \cite{Weiße2006a,Napoli2016}; 
the Chebyshev polynomials $T_i$ form a complete basis for square integrable functions 
which take values in the range $[-1,1]$
and a linear combination of them can be chosen to approximate the eigenspectrum of a 
tight-binding hamiltonian after a simple rescaling to ensure all eigenvalues lie in $[-1,1]$.

An additional advantage of the method is that it can be formulated into a code 
with excellent parallel efficiency, especially compared to DFT super-cell calculations. 
This is a consequence of the fact that to obtain the global operator requires 
a large number of independent computations of the local operator in different 
configurations that can be run in parallel (we use MVAPICH 2.2b). 
Since each local operator is computed using only sparse matrix-vector operations, 
a second layer of parallelization can be added by using multi-threaded implementations 
of highly optimized matrix-vector operator subroutines, which further enhances efficiency (we use Intel MKL 11.0).

\section{\label{sec:level3}BILAYER GRAPHENE} Twisted bilayer graphene (tBLG) provides an excellent 
candidate for a test of our method, 
since it has been well characterized by many experimental works and
analytical theory~\cite{Bistritzer2011,San-Jose2012}. 
To compute the EDoS of tBLG we used a two-band 
model that describes the $\pi$ bonding 
and antibonding combinations of $p_z$ 
orbitals associated with the two-atom basis of the honeycomb lattice; 
the tight-binding hamiltonian is derived from first-principles calculations 
with the use of Wannier orbitals 
and involves no adjustable parameters, other than the 
range of hopping matrix elements~\cite{Fang2016}. 

The main feature of twisted bilayer graphene is the presence of van Hove singularities 
(VHS) above and below the Fermi energy. 
The origin of these VHS can be best understood by considering the low-energy band-structure of tBLG as consisting of four Dirac cones at the valleys $K_l$ and $K'_l$, 
where $l = (1,2)$ labels the layers. 
At $\theta = 0^{\circ}$ twist, $K_1$ and $K_2$ are at the same point in momentum-space. 
For $\theta > 0^{\circ}$, the Dirac cones move away from one another in momentum-space, 
and a partial band-gap opening occurs where the cones now overlap. 
These hybridizations at the overlap of the Dirac cones produce the VHS \cite{CastroNeto2009}, 
which have already been investigated by experimental STM measurements~\cite{Li2009,Luican2011,Wong2015,Yin2015}.

\begin{figure}
			\centering
			\includegraphics[width=8.6 cm]{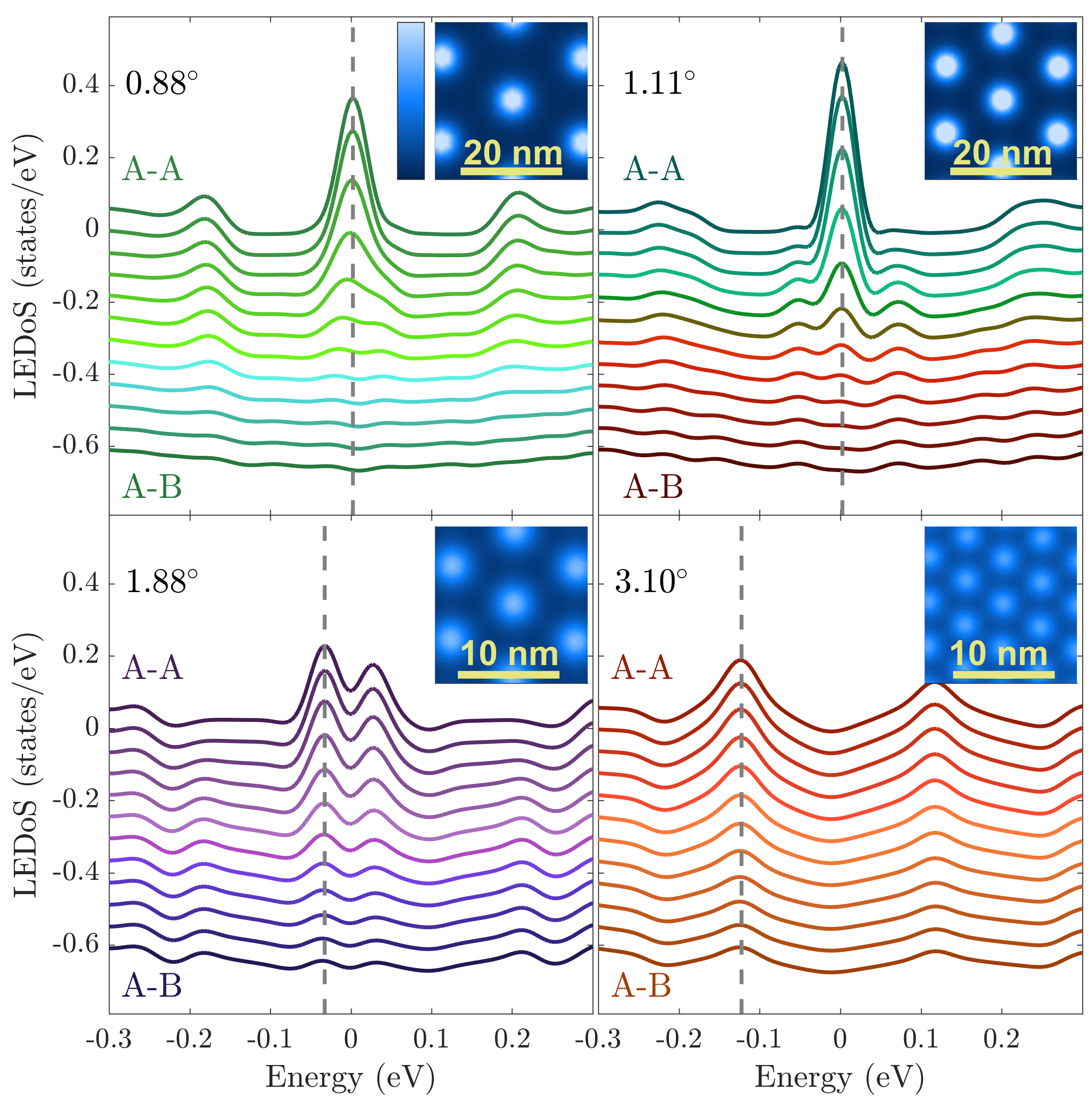}
			\caption{Simulated local electronic density of states (LEDoS) 
at four different angles of twisted bilayer graphene. Each line corresponds 
to a different real-space configuration along the line connecting AA to AB stacking. 
The insets show a real-space image of the density of states in the bilayer system at the
energy value identified by a dashed line. 
The figure was constructed to facilitate 
comparison with experiment \cite{Yin2015}, which 
shows excellent agreement. 
}
			\label{fig:tblg_config_4plot}
		\end{figure}
		
As a first test of the method, in Fig. \ref{fig:tblg_config_4plot}
we compare the spatial dependence of tBLG 
at four twist angles to experimental results~\cite{Yin2015}.
This is possible because sampling shifts over the diagonal of one layer's unit-cell 
is the same as moving linearly from an AA to AB type stacking in the real-space moir\'{e} 
pattern.  For these calculations we use a disk cut-off radius 
$R =  500$ \r{A} which contains $591,344$ atoms.
The simulated features of the VHS for the four selected angles are identical to those from experiment~\cite{Yin2015}, but the scaling between the VHS feature and the 
background graphene DoS are different between theory and experiment. 
This can be partly explained by the fact that in STM measurements states with lower 
in-plane momentum ${\bf k}$ have shorter decay lengths~\cite{Stroscio1986,Huang2015}. 
Our method gives the DoS independent of the momentum of electronic states
that contribute to it, so it is expected that the VHS will be less pronounced 
in experiment. 
		
\begin{figure*}
			\centering
			\includegraphics[width=17.2 cm]{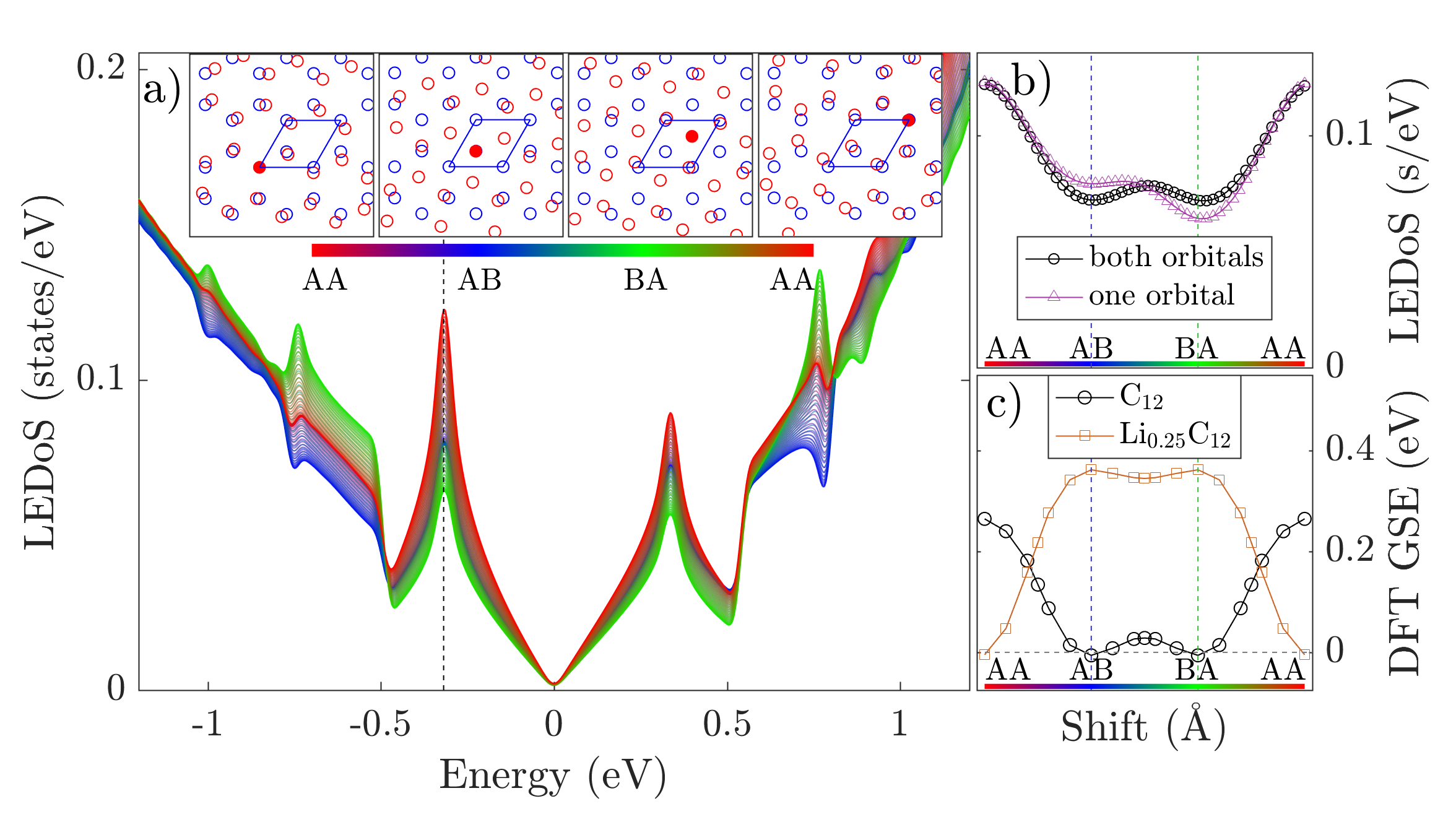}
			\caption{Local Electronic Density of States as a function of shift distance across the unit cell diagonal:
(a) Scan of a single orbital's LEDoS with the coloring corresponding to distance across the diagonal. The insets show the real-space configurations 
$\omega$ for three types of stacking with the atoms in each layer 
represented by different color circles (red and blue for top and bottom). 
The unit-cell of the bottom layer is outlined in blue and the 
shifted orbital is highlighted as a filled red dot. 
(b) LEDoS at the selected VHS peak as a function of shift  
for one orbital (triangles) and for the average of both orbitals (circles). 
The peak varies smoothly with shift and has critical points at the three special 
stacking configurations. 
(c) Ground state energy of $\theta = 0^\circ$ twist-angle 
with and without Li-ion doping (see text for details).
}
			\label{fig:tblg_lc_combined}
		\end{figure*}

Next, we sample the configuration space 
$\Omega$ for a fixed twist angle of $\theta = 5.73^{\circ}$ (0.1 radians) 
for 100 configurations along the diagonal of the unit-cell, see Fig. \ref{fig:tblg_lc_combined}. 
The LEDoS varies smoothly as a function of $\omega$, with the only regions of
significant configuration dependence being those near the VHS. 
The AB (BA) type stacking has much lower DoS at the VHS than any other stacking configuration. Since we fit the DoS to a smooth polynomial, 
the divergent nature of the DoS at the VHS is only partially recovered. 
We can still compare the intensity of the VHS by examining its spectral weight.
In Fig. \ref{fig:tblg_lc_combined} (c) we plot the DFT ground-state energy
calculations for non-twisted bilayer graphene over the same range of 
relative shifts.
There are interesting similarities between the VHS LEDoS and the ground-state energy, 
namely, the LEDoS at the VHS has the same dependence on relative shift as the
energy.
An important question is: can one controllably induce a relative twist between two
graphene layers in samples of macroscopic size?  We 
suggest that use of intercalants may facilitate this process.  In particular, 
Li-ions are know to be easily intercalated between graphene layers, with 
both insertion and removal being fast processes.  Inspired by this observation, 
we have also calculated the ground-state energy 
as a function of relative shift for a graphene bilayer 
including Li-ion intercalation. In the fully lithiated structure, 
the relative stability of the AB and AA stacking is inverted, suggesting that 
Li-ion intercalation may indeed act as a way to facilitate
changes in the relative twist-angle even for macroscopic samples. 
		
\begin{figure*}
			\centering
			\includegraphics[width=17.2 cm]{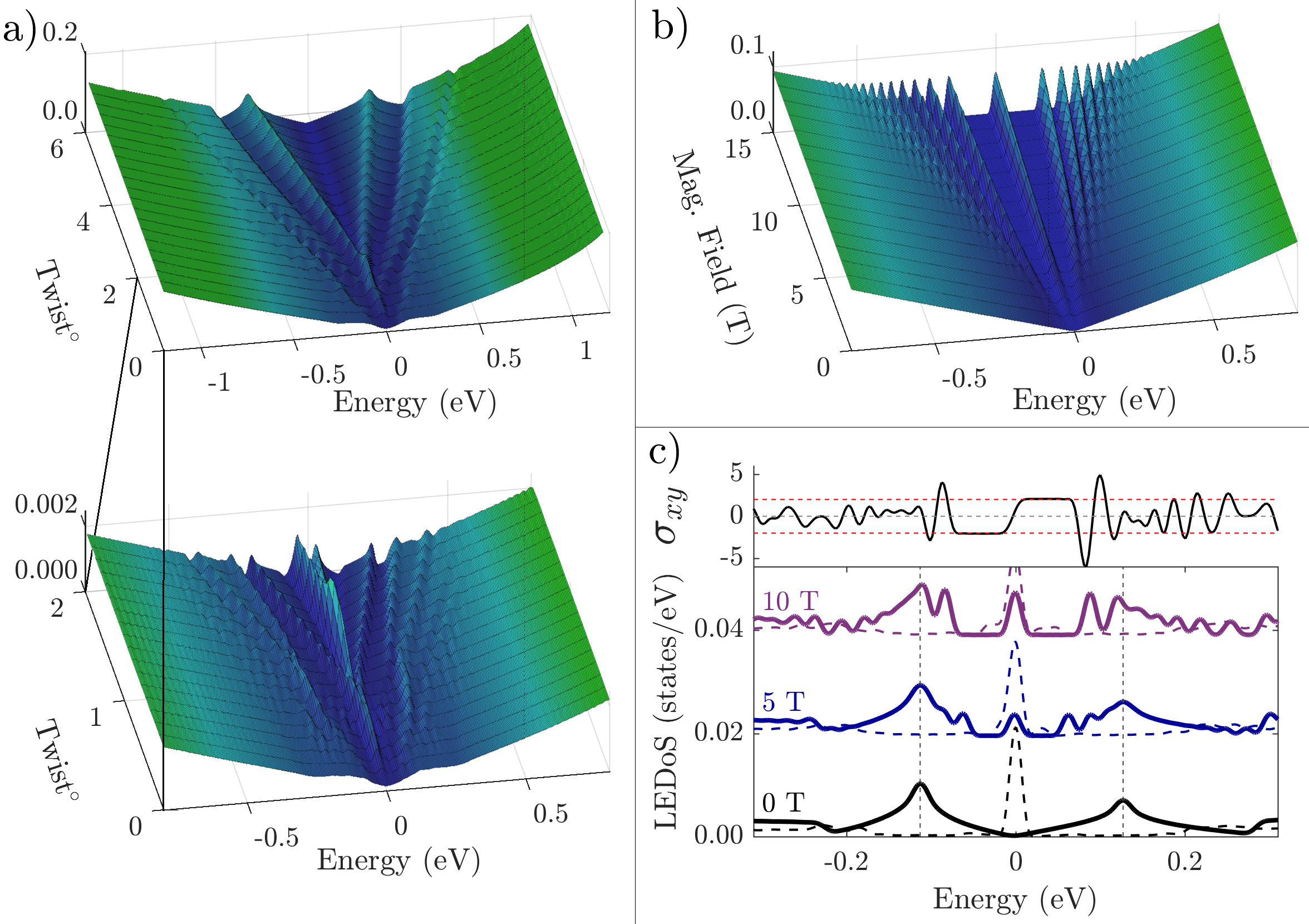}
			\caption{
				(a) Average EDoS as a function of twist angle for tBLG.  
				(b) Average EDoS for monolayer graphene in the presence of varying magnetic field.
				(c) LEDoS for AA stacked tBLG with $3.1^{\circ}$ (solid line) and $1.1^{\circ}$ (dashed line) twist-angle at different values of the magnetic field
and Hall conductivity $\sigma_{xy}$ in units of $e^2/h$, with the horizontal red 
dashed lines at $\pm 2 e^2/h$.
}
			\label{fig:bfield_and_twist_plot}
		\end{figure*}

In Fig. \ref{fig:bfield_and_twist_plot}(a)
we plot the angle-dependent EDoS for tBLG.  
The first, second, and third VHS are visible in the low-angle regime and they move away from the Fermi level linearly with twist angle. 
At the VHS, we find that the real-space local DoS is highly localized at the AA stacking sites as in Fig. \ref{fig:tblg_config_4plot}, in agreement with experimental STM  results \cite{Wong2015, Brihuega2012}. 
It is easy to identify in Fig. \ref{fig:bfield_and_twist_plot}(a) the first and second 
``magic angles'' of tBLG (near $1.1^\circ$ and $0.5^\circ$, respectively), 
explained by band flattening near the Fermi level~\cite{Bistritzer2011, San-Jose2012}. 
In Fig. \ref{fig:bfield_and_twist_plot}(b) we plot the calculated EDoS of 
monolayer graphene in the presence of out-of-plane magnetic field. 
The Landau levels (LL's) in the monolayer and the VHS in the twisted bilayer both represent tunable, localized electronic states, supporting the interpretation of the interlayer 
interaction in a twisted heterostructure as a non-abelian gauge field~\cite{San-Jose2012, Yin2015}. These calculations allow a very robust determination of the monolayer's Fermi velocity without a band-structure calculation, using the low-energy model for the LL's~\cite{CastroNeto2009}
		\begin{equation}
		E(n) = \pm \mathit{v}_F \sqrt{2 e B N}
		\end{equation}
with the result for the Fermi velocity $\mathit{v}_F = 1.2 \times 10^6$ m/s. 
Finally, we test the interaction between twist and magnetic field in 
Fig. \ref{fig:bfield_and_twist_plot}(c): at the AA stacking with a $3.1^\circ$ twist 
there are many clear LL's and at a field of 5 T, the peak of the VHS is significantly altered
relative to its zero-field shape. At a twist of $1.1^\circ$, the magnetic field dependence
of the peak is not visible. These results are in good agreement with experimental STM measurements~\cite{Luican2011, Yin2015}.
		
The St\"{r}eda formula \cite{Streda1982} relates the fluctuations in the integrated electronic density $n$ under small changes in the magnetic field strength $B$ to the Hall conductance
$\sigma_{xy}$,  
while the energy $E$ of the system energy is in an energy-gapped region, $E \in E_g$:
\begin{equation}
\sigma_{xy} = e \frac{\partial n(E)}{\partial B} \bigg|_{E \in E_g}
\end{equation}
Averaging over 100 configurations of the LEDoS on both layers
with a 750 \r{A} cut-off radius ($1,330,550$ atoms), gives values for 
$\sigma_{xy}$ that jump from $-2$ to $+2$ in units of  $e^2/h$ 
across the central LL in the $3.1^\circ$ simulation. 
This change of $+4 e^2/h$, before taking into account spin,  
corresponds to the four-fold degeneracy for the $N = 0$ 
LL of bilayer graphene, with the four states originating from the monolayer's $K, K'$ 
valley degeneracy (factor of 2) and the two sheets (another factor of 2). 
A change of $+8 e^2/h$ is observed in experiment for tBLG, 
which is in agreement with our results when we take into account spin degeneracy~\cite{Cao2016}. 
If only the AA configuration is used in the calculation we do not obtain good 
quantization of $\sigma_{xy}$. 
Just like integrating over the entire Brillouin Zone when computing in momentum space, 
integrating over the entire configuration space $\Omega$ is required in the case of tBLG. 
This allows us to compute the Chern number for the wavefunctions in the gapped region 
by taking the difference in $\sigma_{xy}$ in units of the conductance quanta ($+4 e^2/h$), which indicates that our method can capture accurately certain topological properties of the electronic band structure.
		
%	\subsection{Twisted MX$_2$ TMDC Bilayers}
	
	\section{\label{sec:level4}BILAYER TMDC}  
	
Unlike bilayer graphene, transition metal dichalcogenides will not be well described by 
low-energy theory due to their large band-gaps (about 2 eV). 
For bilayers of TMDCs we use an 11 band model, consisting of 5 $d$ orbitals 
on the transition metal atom and 3 $p$ orbitals on each of the two chalcogen atoms~\cite{Fang2015}. 
The interlayer interaction is modeled only between the chalcogen atoms closest 
to the bilayer interface. Here we present results for MoS$_2$, 
whose model hamiltonian includes GW corrections for more accurate 
representation of the electronic structure. 
Since we are mainly interested in studying twist-angle dependent effects, 
we will neglect spin polarization, 
but an \textit{ab-initio} model with spin-orbit coupling can be easily
substituted if such effects are important.
	
\begin{figure}
		\centering
		\includegraphics[width=8.6 cm]{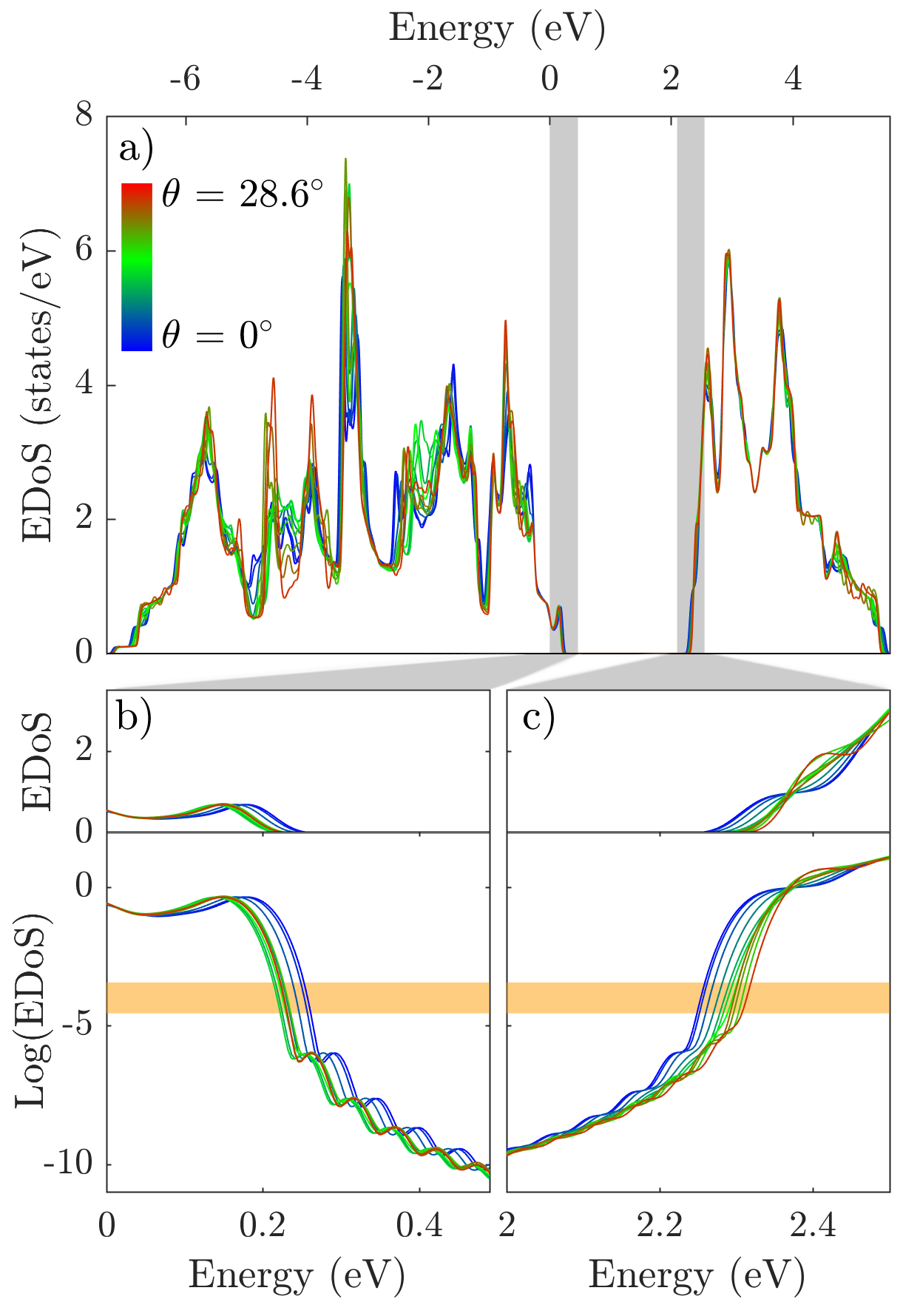}
		\caption{
(a) EDoS for twisted bilayer MoS$_2$ from $0^{\circ}$ (blue) to $28.65^{\circ}$ (red) twist angle. 
(b) and (c) EDoS near the valence and conduction band extrema, with the 
logarithmic scales showing the changes in greater detail. }
		\label{fig:mos2_anglesweep_bandgap}
	\end{figure}

Some twist-angle dependent features were seen in the LEDoS for both WSe$_2$ and MoS$_2$, but most were not near the conduction or valence band-edges. 
The twist-angle dependence of the density of states for bilayer MoS$_2$ is shown in Fig. \ref{fig:mos2_anglesweep_bandgap}(a).
These calculations were performed with a 300 \r{A} cut-off radius ($193,700$ atoms) and averaged over 100 configurations.
There are significant changes in the EDoS deep into the valence band
(more than 2 eV below the maximum), but it is difficult to probe this 
region experimentally. They could be observed as interesting properties for high-frequency conductivity or optical activity.  

Instead, we focus on the valence and conduction band edges.
The band-gap is a twist-angle dependent feature: it increases by 
76 meV (a $\sim 4$\% change) 
going from $0^{\circ}$ to $28.6^{\circ}$ twist-angle.
The regions near the valence and conduction band extrema are shown in great detail in 
Fig. \ref{fig:mos2_anglesweep_bandgap}(b) and (c), with the logarithmic scale 
showing the changes more clearly. These plots also show the good numerical convergence 
of the EDoS in our model, with noticeable numeric error only occurring when the EDoS is smaller than $10^{-5}$ states per eV. This error, reminiscent of Gibbs oscillations \cite{Weiße2006a}, is likely an artifact of the KPM attempting to fit a smooth function to a band-edge in the eigenvalue spectrum. We thus take a region about $10^{-4}$ states per eV to compare changes in the band-gap (plotted in orange).
Our model does not take into account changes in the distance between the two layers as a function of twist-angle, which could give additional dependence of the band-gap and can be 
incorporated as a dependence of the tight-binding hopping matrix elements on twist 
angle and distance. 
	
\section{\label{sec:level5}CONCLUSION} 
	
We have introduced a new method for parameter-free computation of electronic properties in incommensurate layered 2D materials with controllable errors.
Although here we have only studied bilayer materials, the method is general and extends to any number of layers and of arbitrary heterostructure composition.
Viewing the problem on the space of configurations, $\Omega$, allows us to fully characterize the properties of incommensurate (aperiodic) systems.
The method allows for the inclusion of external fields and other sources of disorder, such as strain or defects.
We present results of applying the method to twisted bilayer graphene and a
representative of the TMDC family of semiconductors.
The method is accurate enough to correctly calculate quantization of 
Hall conductivity in tBLG in the presence of magnetic fields, 
and reproduces the correct Chern number for the $N = 0$ Landau Level.
It also predicts that bilayer TMDC's have a twist-dependent band-gap.
The method is a promising candidate for the targeted design of 
electronic properties in layered heterostructures.
	
\begin{acknowledgements}
We acknowledge S. Shirodkar for providing the Li-ion itercalated graphene calculations shown in Fig. \ref{fig:tblg_lc_combined}(c) and B.I. Halperin and D. Huang 
for helpful discussions. The computations in this paper were run on the Odyssey cluster supported by the FAS Division of Science, Research Computing Group at Harvard University.
This work was supported by the ARO MURI Award No. W911NF-14-0247. 
SF is supported by the STC Center for Integrated Quantum Materials, NSF Grant No. DMR-1231319.
\end{acknowledgements}

	\bibliographystyle{apsrev4-1}
	\bibliography{twistronics_papers}

\begin{thebibliography}{32}%
\makeatletter
\providecommand \@ifxundefined [1]{%
 \@ifx{#1\undefined}
}%
\providecommand \@ifnum [1]{%
 \ifnum #1\expandafter \@firstoftwo
 \else \expandafter \@secondoftwo
 \fi
}%
\providecommand \@ifx [1]{%
 \ifx #1\expandafter \@firstoftwo
 \else \expandafter \@secondoftwo
 \fi
}%
\providecommand \natexlab [1]{#1}%
\providecommand \enquote  [1]{``#1''}%
\providecommand \bibnamefont  [1]{#1}%
\providecommand \bibfnamefont [1]{#1}%
\providecommand \citenamefont [1]{#1}%
\providecommand \href@noop [0]{\@secondoftwo}%
\providecommand \href [0]{\begingroup \@sanitize@url \@href}%
\providecommand \@href[1]{\@@startlink{#1}\@@href}%
\providecommand \@@href[1]{\endgroup#1\@@endlink}%
\providecommand \@sanitize@url [0]{\catcode `\\12\catcode `\$12\catcode
  `\&12\catcode `\#12\catcode `\^12\catcode `\_12\catcode `\%12\relax}%
\providecommand \@@startlink[1]{}%
\providecommand \@@endlink[0]{}%
\providecommand \url  [0]{\begingroup\@sanitize@url \@url }%
\providecommand \@url [1]{\endgroup\@href {#1}{\urlprefix }}%
\providecommand \urlprefix  [0]{URL }%
\providecommand \Eprint [0]{\href }%
\providecommand \doibase [0]{http://dx.doi.org/}%
\providecommand \selectlanguage [0]{\@gobble}%
\providecommand \bibinfo  [0]{\@secondoftwo}%
\providecommand \bibfield  [0]{\@secondoftwo}%
\providecommand \translation [1]{[#1]}%
\providecommand \BibitemOpen [0]{}%
\providecommand \bibitemStop [0]{}%
\providecommand \bibitemNoStop [0]{.\EOS\space}%
\providecommand \EOS [0]{\spacefactor3000\relax}%
\providecommand \BibitemShut  [1]{\csname bibitem#1\endcsname}%
\let\auto@bib@innerbib\@empty
%</preamble>
\bibitem [{\citenamefont {Novoselov}\ \emph {et~al.}(2004)\citenamefont
  {Novoselov}, \citenamefont {Geim}, \citenamefont {Morozov}, \citenamefont
  {Jiang}, \citenamefont {Zhang}, \citenamefont {Dubonos}, \citenamefont
  {Grigorieva},\ and\ \citenamefont {Firsov}}]{Novoselov2004}%
  \BibitemOpen
  \bibfield  {author} {\bibinfo {author} {\bibfnamefont {K.~S.}\ \bibnamefont
  {Novoselov}}, \bibinfo {author} {\bibfnamefont {A.~K.}\ \bibnamefont {Geim}},
  \bibinfo {author} {\bibfnamefont {S.~V.}\ \bibnamefont {Morozov}}, \bibinfo
  {author} {\bibfnamefont {D.}~\bibnamefont {Jiang}}, \bibinfo {author}
  {\bibfnamefont {Y.}~\bibnamefont {Zhang}}, \bibinfo {author} {\bibfnamefont
  {S.~V.}\ \bibnamefont {Dubonos}}, \bibinfo {author} {\bibfnamefont {I.~V.}\
  \bibnamefont {Grigorieva}}, \ and\ \bibinfo {author} {\bibfnamefont {A.~A.}\
  \bibnamefont {Firsov}},\ }\href {\doibase 10.1126/science.1102896} {\bibfield
   {journal} {\bibinfo  {journal} {Science (New York, N.Y.)}\ }\textbf
  {\bibinfo {volume} {306}},\ \bibinfo {pages} {666} (\bibinfo {year}
  {2004})}\BibitemShut {NoStop}%
\bibitem [{\citenamefont {Ayari}\ \emph {et~al.}(2007)\citenamefont {Ayari},
  \citenamefont {Cobas}, \citenamefont {Ogundadegbe},\ and\ \citenamefont
  {Fuhrer}}]{Ayari2007}%
  \BibitemOpen
  \bibfield  {author} {\bibinfo {author} {\bibfnamefont {A.}~\bibnamefont
  {Ayari}}, \bibinfo {author} {\bibfnamefont {E.}~\bibnamefont {Cobas}},
  \bibinfo {author} {\bibfnamefont {O.}~\bibnamefont {Ogundadegbe}}, \ and\
  \bibinfo {author} {\bibfnamefont {M.~S.}\ \bibnamefont {Fuhrer}},\ }\href
  {\doibase 10.1063/1.2407388} {\bibfield  {journal} {\bibinfo  {journal}
  {Journal of Applied Physics}\ }\textbf {\bibinfo {volume} {101}},\ \bibinfo
  {pages} {014507} (\bibinfo {year} {2007})}\BibitemShut {NoStop}%
\bibitem [{\citenamefont {Dean}\ \emph {et~al.}(2010)\citenamefont {Dean},
  \citenamefont {Young}, \citenamefont {Meric}, \citenamefont {Lee},
  \citenamefont {Wang}, \citenamefont {Sorgenfrei}, \citenamefont {Watanabe},
  \citenamefont {Taniguchi}, \citenamefont {Kim}, \citenamefont {Shepard},\
  and\ \citenamefont {Hone}}]{Dean2010}%
  \BibitemOpen
  \bibfield  {author} {\bibinfo {author} {\bibfnamefont {C.~R.}\ \bibnamefont
  {Dean}}, \bibinfo {author} {\bibfnamefont {a.~F.}\ \bibnamefont {Young}},
  \bibinfo {author} {\bibfnamefont {I.}~\bibnamefont {Meric}}, \bibinfo
  {author} {\bibfnamefont {C.}~\bibnamefont {Lee}}, \bibinfo {author}
  {\bibfnamefont {L.}~\bibnamefont {Wang}}, \bibinfo {author} {\bibfnamefont
  {S.}~\bibnamefont {Sorgenfrei}}, \bibinfo {author} {\bibfnamefont
  {K.}~\bibnamefont {Watanabe}}, \bibinfo {author} {\bibfnamefont
  {T.}~\bibnamefont {Taniguchi}}, \bibinfo {author} {\bibfnamefont
  {P.}~\bibnamefont {Kim}}, \bibinfo {author} {\bibfnamefont {K.~L.}\
  \bibnamefont {Shepard}}, \ and\ \bibinfo {author} {\bibfnamefont
  {J.}~\bibnamefont {Hone}},\ }\href {\doibase 10.1038/nnano.2010.172}
  {\bibfield  {journal} {\bibinfo  {journal} {Nature nanotechnology}\ }\textbf
  {\bibinfo {volume} {5}},\ \bibinfo {pages} {722} (\bibinfo {year}
  {2010})}\BibitemShut {NoStop}%
\bibitem [{\citenamefont {Mak}\ \emph {et~al.}(2010)\citenamefont {Mak},
  \citenamefont {Lee}, \citenamefont {Hone}, \citenamefont {Shan},\ and\
  \citenamefont {Heinz}}]{Mak2010}%
  \BibitemOpen
  \bibfield  {author} {\bibinfo {author} {\bibfnamefont {K.~F.}\ \bibnamefont
  {Mak}}, \bibinfo {author} {\bibfnamefont {C.}~\bibnamefont {Lee}}, \bibinfo
  {author} {\bibfnamefont {J.}~\bibnamefont {Hone}}, \bibinfo {author}
  {\bibfnamefont {J.}~\bibnamefont {Shan}}, \ and\ \bibinfo {author}
  {\bibfnamefont {T.~F.}\ \bibnamefont {Heinz}},\ }\href {\doibase
  10.1103/PhysRevLett.105.136805} {\bibfield  {journal} {\bibinfo  {journal}
  {Physical review letters}\ }\textbf {\bibinfo {volume} {105}},\ \bibinfo
  {pages} {136805} (\bibinfo {year} {2010})}\BibitemShut {NoStop}%
\bibitem [{\citenamefont {Radisavljevic}\ \emph {et~al.}(2011)\citenamefont
  {Radisavljevic}, \citenamefont {Radenovic}, \citenamefont {Brivio},
  \citenamefont {Giacometti},\ and\ \citenamefont {Kis}}]{Radisavljevic2011}%
  \BibitemOpen
  \bibfield  {author} {\bibinfo {author} {\bibfnamefont {B.}~\bibnamefont
  {Radisavljevic}}, \bibinfo {author} {\bibfnamefont {A.}~\bibnamefont
  {Radenovic}}, \bibinfo {author} {\bibfnamefont {J.}~\bibnamefont {Brivio}},
  \bibinfo {author} {\bibfnamefont {V.}~\bibnamefont {Giacometti}}, \ and\
  \bibinfo {author} {\bibfnamefont {A.}~\bibnamefont {Kis}},\ }\href {\doibase
  10.1038/nnano.2010.279} {\bibfield  {journal} {\bibinfo  {journal} {Nature
  nanotechnology}\ }\textbf {\bibinfo {volume} {6}},\ \bibinfo {pages} {147}
  (\bibinfo {year} {2011})}\BibitemShut {NoStop}%
\bibitem [{\citenamefont {De}\ \emph {et~al.}(2013)\citenamefont {De},
  \citenamefont {Manongdo}, \citenamefont {See}, \citenamefont {Zhang},
  \citenamefont {Guloy},\ and\ \citenamefont {Peng}}]{De2013}%
  \BibitemOpen
  \bibfield  {author} {\bibinfo {author} {\bibfnamefont {D.}~\bibnamefont
  {De}}, \bibinfo {author} {\bibfnamefont {J.}~\bibnamefont {Manongdo}},
  \bibinfo {author} {\bibfnamefont {S.}~\bibnamefont {See}}, \bibinfo {author}
  {\bibfnamefont {V.}~\bibnamefont {Zhang}}, \bibinfo {author} {\bibfnamefont
  {A.}~\bibnamefont {Guloy}}, \ and\ \bibinfo {author} {\bibfnamefont
  {H.}~\bibnamefont {Peng}},\ }\href {\doibase 10.1088/0957-4484/24/2/025202}
  {\bibfield  {journal} {\bibinfo  {journal} {Nanotechnology}\ }\textbf
  {\bibinfo {volume} {24}},\ \bibinfo {pages} {025202} (\bibinfo {year}
  {2013})}\BibitemShut {NoStop}%
\bibitem [{\citenamefont {Geim}\ and\ \citenamefont
  {Grigorieva}(2013)}]{Geim2013}%
  \BibitemOpen
  \bibfield  {author} {\bibinfo {author} {\bibfnamefont {A.~K.}\ \bibnamefont
  {Geim}}\ and\ \bibinfo {author} {\bibfnamefont {I.~V.}\ \bibnamefont
  {Grigorieva}},\ }\href {\doibase 10.1038/nature12385} {\bibfield  {journal}
  {\bibinfo  {journal} {Nature}\ }\textbf {\bibinfo {volume} {499}},\ \bibinfo
  {pages} {419} (\bibinfo {year} {2013})}\BibitemShut {NoStop}%
\bibitem [{\citenamefont {Cao}\ \emph {et~al.}(2016)\citenamefont {Cao},
  \citenamefont {Luo}, \citenamefont {Fatemi}, \citenamefont {Fang},
  \citenamefont {Sanchez-Yamagishi}, \citenamefont {Watanabe}, \citenamefont
  {Taniguchi}, \citenamefont {Kaxiras},\ and\ \citenamefont
  {Jarillo-Herrero}}]{Cao2016}%
  \BibitemOpen
  \bibfield  {author} {\bibinfo {author} {\bibfnamefont {Y.}~\bibnamefont
  {Cao}}, \bibinfo {author} {\bibfnamefont {J.~Y.}\ \bibnamefont {Luo}},
  \bibinfo {author} {\bibfnamefont {V.}~\bibnamefont {Fatemi}}, \bibinfo
  {author} {\bibfnamefont {S.}~\bibnamefont {Fang}}, \bibinfo {author}
  {\bibfnamefont {J.~D.}\ \bibnamefont {Sanchez-Yamagishi}}, \bibinfo {author}
  {\bibfnamefont {K.}~\bibnamefont {Watanabe}}, \bibinfo {author}
  {\bibfnamefont {T.}~\bibnamefont {Taniguchi}}, \bibinfo {author}
  {\bibfnamefont {E.}~\bibnamefont {Kaxiras}}, \ and\ \bibinfo {author}
  {\bibfnamefont {P.}~\bibnamefont {Jarillo-Herrero}},\ }\href {\doibase
  10.1103/PhysRevLett.117.116804} {\bibfield  {journal} {\bibinfo  {journal}
  {Physical Review Letters}\ }\textbf {\bibinfo {volume} {117}},\ \bibinfo
  {pages} {116804} (\bibinfo {year} {2016})}\BibitemShut {NoStop}%
\bibitem [{\citenamefont {Koren}\ \emph {et~al.}(2016)\citenamefont {Koren},
  \citenamefont {Leven}, \citenamefont {L\"{o}rtscher}, \citenamefont {Knoll},
  \citenamefont {Hod},\ and\ \citenamefont {Duerig}}]{Koren2016}%
  \BibitemOpen
  \bibfield  {author} {\bibinfo {author} {\bibfnamefont {E.}~\bibnamefont
  {Koren}}, \bibinfo {author} {\bibfnamefont {I.}~\bibnamefont {Leven}},
  \bibinfo {author} {\bibfnamefont {E.}~\bibnamefont {L\"{o}rtscher}}, \bibinfo
  {author} {\bibfnamefont {A.}~\bibnamefont {Knoll}}, \bibinfo {author}
  {\bibfnamefont {O.}~\bibnamefont {Hod}}, \ and\ \bibinfo {author}
  {\bibfnamefont {U.}~\bibnamefont {Duerig}},\ }\href {\doibase
  10.1038/nnano.2016.85} {\bibfield  {journal} {\bibinfo  {journal} {Nature
  nanotechnology}\ }\textbf {\bibinfo {volume} {11}},\ \bibinfo {pages} {752}
  (\bibinfo {year} {2016})}\BibitemShut {NoStop}%
\bibitem [{\citenamefont {Rozhkov}\ \emph {et~al.}(2016)\citenamefont
  {Rozhkov}, \citenamefont {Sboychakov}, \citenamefont {Rakhmanov},\ and\
  \citenamefont {Nori}}]{Rozhkov2016}%
  \BibitemOpen
  \bibfield  {author} {\bibinfo {author} {\bibfnamefont {A.}~\bibnamefont
  {Rozhkov}}, \bibinfo {author} {\bibfnamefont {A.}~\bibnamefont {Sboychakov}},
  \bibinfo {author} {\bibfnamefont {A.}~\bibnamefont {Rakhmanov}}, \ and\
  \bibinfo {author} {\bibfnamefont {F.}~\bibnamefont {Nori}},\ }\href {\doibase
  10.1016/j.physrep.2016.07.003} {\bibfield  {journal} {\bibinfo  {journal}
  {Physics Reports}\ }\textbf {\bibinfo {volume} {648}},\ \bibinfo {pages} {1}
  (\bibinfo {year} {2016})}\BibitemShut {NoStop}%
\bibitem [{\citenamefont {Uchida}\ \emph {et~al.}(2014)\citenamefont {Uchida},
  \citenamefont {Furuya}, \citenamefont {Iwata},\ and\ \citenamefont
  {Oshiyama}}]{Uchida2014}%
  \BibitemOpen
  \bibfield  {author} {\bibinfo {author} {\bibfnamefont {K.}~\bibnamefont
  {Uchida}}, \bibinfo {author} {\bibfnamefont {S.}~\bibnamefont {Furuya}},
  \bibinfo {author} {\bibfnamefont {J.-I.}\ \bibnamefont {Iwata}}, \ and\
  \bibinfo {author} {\bibfnamefont {A.}~\bibnamefont {Oshiyama}},\ }\href
  {\doibase 10.1103/PhysRevB.90.155451} {\bibfield  {journal} {\bibinfo
  {journal} {Physical Review B}\ }\textbf {\bibinfo {volume} {90}},\ \bibinfo
  {pages} {155451} (\bibinfo {year} {2014})}\BibitemShut {NoStop}%
\bibitem [{\citenamefont {Woods}\ \emph {et~al.}(2014)\citenamefont {Woods},
  \citenamefont {Britnell}, \citenamefont {Eckmann}, \citenamefont {Ma},
  \citenamefont {Lu}, \citenamefont {Guo}, \citenamefont {Lin}, \citenamefont
  {Yu}, \citenamefont {Cao}, \citenamefont {Gorbachev}, \citenamefont
  {Kretinin}, \citenamefont {Park}, \citenamefont {Ponomarenko}, \citenamefont
  {Katsnelson}, \citenamefont {Gornostyrev}, \citenamefont {Watanabe},
  \citenamefont {Taniguchi}, \citenamefont {Casiraghi}, \citenamefont {Gao},
  \citenamefont {Geim},\ and\ \citenamefont {Novoselov}}]{Woods2014}%
  \BibitemOpen
  \bibfield  {author} {\bibinfo {author} {\bibfnamefont {C.~R.}\ \bibnamefont
  {Woods}}, \bibinfo {author} {\bibfnamefont {L.}~\bibnamefont {Britnell}},
  \bibinfo {author} {\bibfnamefont {A.}~\bibnamefont {Eckmann}}, \bibinfo
  {author} {\bibfnamefont {R.~S.}\ \bibnamefont {Ma}}, \bibinfo {author}
  {\bibfnamefont {J.~C.}\ \bibnamefont {Lu}}, \bibinfo {author} {\bibfnamefont
  {H.~M.}\ \bibnamefont {Guo}}, \bibinfo {author} {\bibfnamefont
  {X.}~\bibnamefont {Lin}}, \bibinfo {author} {\bibfnamefont {G.~L.}\
  \bibnamefont {Yu}}, \bibinfo {author} {\bibfnamefont {Y.}~\bibnamefont
  {Cao}}, \bibinfo {author} {\bibfnamefont {R.~V.}\ \bibnamefont {Gorbachev}},
  \bibinfo {author} {\bibfnamefont {A.~V.}\ \bibnamefont {Kretinin}}, \bibinfo
  {author} {\bibfnamefont {J.}~\bibnamefont {Park}}, \bibinfo {author}
  {\bibfnamefont {L.~A.}\ \bibnamefont {Ponomarenko}}, \bibinfo {author}
  {\bibfnamefont {M.~I.}\ \bibnamefont {Katsnelson}}, \bibinfo {author}
  {\bibfnamefont {Y.~N.}\ \bibnamefont {Gornostyrev}}, \bibinfo {author}
  {\bibfnamefont {K.}~\bibnamefont {Watanabe}}, \bibinfo {author}
  {\bibfnamefont {T.}~\bibnamefont {Taniguchi}}, \bibinfo {author}
  {\bibfnamefont {C.}~\bibnamefont {Casiraghi}}, \bibinfo {author}
  {\bibfnamefont {H.-J.}\ \bibnamefont {Gao}}, \bibinfo {author} {\bibfnamefont
  {A.~K.}\ \bibnamefont {Geim}}, \ and\ \bibinfo {author} {\bibfnamefont
  {K.~S.}\ \bibnamefont {Novoselov}},\ }\href {\doibase 10.1038/nphys2954}
  {\bibfield  {journal} {\bibinfo  {journal} {Nature Physics}\ }\textbf
  {\bibinfo {volume} {10}},\ \bibinfo {pages} {451} (\bibinfo {year}
  {2014})}\BibitemShut {NoStop}%
\bibitem [{\citenamefont {Ebnonnasir}\ \emph {et~al.}(2014)\citenamefont
  {Ebnonnasir}, \citenamefont {Narayanan}, \citenamefont {Kodambaka},\ and\
  \citenamefont {Ciobanu}}]{Ebnonnasir2014}%
  \BibitemOpen
  \bibfield  {author} {\bibinfo {author} {\bibfnamefont {A.}~\bibnamefont
  {Ebnonnasir}}, \bibinfo {author} {\bibfnamefont {B.}~\bibnamefont
  {Narayanan}}, \bibinfo {author} {\bibfnamefont {S.}~\bibnamefont
  {Kodambaka}}, \ and\ \bibinfo {author} {\bibfnamefont {C.~V.}\ \bibnamefont
  {Ciobanu}},\ }\href {\doibase 10.1063/1.4891430} {\bibfield  {journal}
  {\bibinfo  {journal} {Applied Physics Letters}\ }\textbf {\bibinfo {volume}
  {105}},\ \bibinfo {pages} {031603} (\bibinfo {year} {2014})}\BibitemShut
  {NoStop}%
\bibitem [{\citenamefont {Bellissard}\ \emph {et~al.}(1994)\citenamefont
  {Bellissard}, \citenamefont {van Elst},\ and\ \citenamefont {{Schulz-
  Baldes}}}]{Bellissard1994}%
  \BibitemOpen
  \bibfield  {author} {\bibinfo {author} {\bibfnamefont {J.}~\bibnamefont
  {Bellissard}}, \bibinfo {author} {\bibfnamefont {A.}~\bibnamefont {van
  Elst}}, \ and\ \bibinfo {author} {\bibfnamefont {H.}~\bibnamefont {{Schulz-
  Baldes}}},\ }\href {\doibase 10.1063/1.530758} {\bibfield  {journal}
  {\bibinfo  {journal} {Journal of Mathematical Physics}\ }\textbf {\bibinfo
  {volume} {35}},\ \bibinfo {pages} {5373} (\bibinfo {year}
  {1994})}\BibitemShut {NoStop}%
\bibitem [{\citenamefont {Prodan}(2012)}]{Prodan2012}%
  \BibitemOpen
  \bibfield  {author} {\bibinfo {author} {\bibfnamefont {E.}~\bibnamefont
  {Prodan}},\ }\href {\doibase 10.1093/amrx/abs017} {\bibfield  {journal}
  {\bibinfo  {journal} {Applied Mathematics Research eXpress}\ }\textbf
  {\bibinfo {volume} {2}},\ \bibinfo {pages} {176} (\bibinfo {year}
  {2012})}\BibitemShut {NoStop}%
\bibitem [{\citenamefont {Chen}\ and\ \citenamefont {Ortner}(2016)}]{Chen2015}%
  \BibitemOpen
  \bibfield  {author} {\bibinfo {author} {\bibfnamefont {H.}~\bibnamefont
  {Chen}}\ and\ \bibinfo {author} {\bibfnamefont {C.}~\bibnamefont {Ortner}},\
  }\href {\doibase 10.1137/15M1022628} {\bibfield  {journal} {\bibinfo
  {journal} {SIAM Multiscale Model. Simul.}\ }\textbf {\bibinfo {volume}
  {14}},\ \bibinfo {pages} {232} (\bibinfo {year} {2016})}\BibitemShut
  {NoStop}%
\bibitem [{\citenamefont {Massatt}\ \emph {et~al.}(2016)\citenamefont
  {Massatt}, \citenamefont {Luskin},\ and\ \citenamefont
  {Ortner}}]{Massatt2016}%
  \BibitemOpen
  \bibfield  {author} {\bibinfo {author} {\bibfnamefont {D.}~\bibnamefont
  {Massatt}}, \bibinfo {author} {\bibfnamefont {M.}~\bibnamefont {Luskin}}, \
  and\ \bibinfo {author} {\bibfnamefont {C.}~\bibnamefont {Ortner}},\ }\href
  {https://arxiv.org/abs/1608.01968} {\bibfield  {journal} {\bibinfo  {journal}
  {(unpublished)}\ ,\ \bibinfo {pages} {1}} (\bibinfo {year} {2016})},\ \Eprint
  {http://arxiv.org/abs/1608.01968} {arXiv:1608.01968} \BibitemShut {NoStop}%
\bibitem [{\citenamefont {Fang}\ and\ \citenamefont
  {Kaxiras}(2016)}]{Fang2016}%
  \BibitemOpen
  \bibfield  {author} {\bibinfo {author} {\bibfnamefont {S.}~\bibnamefont
  {Fang}}\ and\ \bibinfo {author} {\bibfnamefont {E.}~\bibnamefont {Kaxiras}},\
  }\href {\doibase 10.1103/PhysRevB.93.235153} {\bibfield  {journal} {\bibinfo
  {journal} {Physical Review B}\ }\textbf {\bibinfo {volume} {93}},\ \bibinfo
  {pages} {235153} (\bibinfo {year} {2016})}\BibitemShut {NoStop}%
\bibitem [{\citenamefont {Fang}\ \emph {et~al.}(2015)\citenamefont {Fang},
  \citenamefont {{Kuate Defo}}, \citenamefont {Shirodkar}, \citenamefont
  {Lieu}, \citenamefont {Tritsaris},\ and\ \citenamefont {Kaxiras}}]{Fang2015}%
  \BibitemOpen
  \bibfield  {author} {\bibinfo {author} {\bibfnamefont {S.}~\bibnamefont
  {Fang}}, \bibinfo {author} {\bibfnamefont {R.}~\bibnamefont {{Kuate Defo}}},
  \bibinfo {author} {\bibfnamefont {S.~N.}\ \bibnamefont {Shirodkar}}, \bibinfo
  {author} {\bibfnamefont {S.}~\bibnamefont {Lieu}}, \bibinfo {author}
  {\bibfnamefont {G.~a.}\ \bibnamefont {Tritsaris}}, \ and\ \bibinfo {author}
  {\bibfnamefont {E.}~\bibnamefont {Kaxiras}},\ }\href {\doibase
  10.1103/PhysRevB.92.205108} {\bibfield  {journal} {\bibinfo  {journal}
  {Physical Review B}\ }\textbf {\bibinfo {volume} {92}},\ \bibinfo {pages}
  {205108} (\bibinfo {year} {2015})}\BibitemShut {NoStop}%
\bibitem [{\citenamefont {Bistritzer}\ and\ \citenamefont
  {MacDonald}(2011)}]{Bistritzer2011}%
  \BibitemOpen
  \bibfield  {author} {\bibinfo {author} {\bibfnamefont {R.}~\bibnamefont
  {Bistritzer}}\ and\ \bibinfo {author} {\bibfnamefont {A.~H.}\ \bibnamefont
  {MacDonald}},\ }\href {\doibase 10.1073/pnas.1108174108} {\bibfield
  {journal} {\bibinfo  {journal} {Proceedings of the National Academy of
  Sciences of the United States of America}\ }\textbf {\bibinfo {volume}
  {108}},\ \bibinfo {pages} {12233} (\bibinfo {year} {2011})}\BibitemShut
  {NoStop}%
\bibitem [{\citenamefont {San-Jose}\ \emph {et~al.}(2012)\citenamefont
  {San-Jose}, \citenamefont {Gonz\'{a}lez},\ and\ \citenamefont
  {Guinea}}]{San-Jose2012}%
  \BibitemOpen
  \bibfield  {author} {\bibinfo {author} {\bibfnamefont {P.}~\bibnamefont
  {San-Jose}}, \bibinfo {author} {\bibfnamefont {J.}~\bibnamefont
  {Gonz\'{a}lez}}, \ and\ \bibinfo {author} {\bibfnamefont {F.}~\bibnamefont
  {Guinea}},\ }\href {\doibase 10.1103/PhysRevLett.108.216802} {\bibfield
  {journal} {\bibinfo  {journal} {Physical review letters}\ }\textbf {\bibinfo
  {volume} {108}},\ \bibinfo {pages} {216802} (\bibinfo {year}
  {2012})}\BibitemShut {NoStop}%
\bibitem [{\citenamefont {{Castro Neto}}\ \emph {et~al.}(2009)\citenamefont
  {{Castro Neto}}, \citenamefont {Guinea}, \citenamefont {Peres}, \citenamefont
  {Novoselov},\ and\ \citenamefont {Geim}}]{CastroNeto2009}%
  \BibitemOpen
  \bibfield  {author} {\bibinfo {author} {\bibfnamefont {a.~H.}\ \bibnamefont
  {{Castro Neto}}}, \bibinfo {author} {\bibfnamefont {F.}~\bibnamefont
  {Guinea}}, \bibinfo {author} {\bibfnamefont {N.~M.~R.}\ \bibnamefont
  {Peres}}, \bibinfo {author} {\bibfnamefont {K.~S.}\ \bibnamefont
  {Novoselov}}, \ and\ \bibinfo {author} {\bibfnamefont {a.~K.}\ \bibnamefont
  {Geim}},\ }\href {\doibase 10.1103/RevModPhys.81.109} {\bibfield  {journal}
  {\bibinfo  {journal} {Reviews of Modern Physics}\ }\textbf {\bibinfo {volume}
  {81}},\ \bibinfo {pages} {109} (\bibinfo {year} {2009})}\BibitemShut
  {NoStop}%
\bibitem [{\citenamefont {Li}\ \emph {et~al.}(2009)\citenamefont {Li},
  \citenamefont {Luican}, \citenamefont {{Lopes dos Santos}}, \citenamefont
  {{Castro Neto}}, \citenamefont {Reina}, \citenamefont {Kong},\ and\
  \citenamefont {Andrei}}]{Li2009}%
  \BibitemOpen
  \bibfield  {author} {\bibinfo {author} {\bibfnamefont {G.}~\bibnamefont
  {Li}}, \bibinfo {author} {\bibfnamefont {A.}~\bibnamefont {Luican}}, \bibinfo
  {author} {\bibfnamefont {J.~M.~B.}\ \bibnamefont {{Lopes dos Santos}}},
  \bibinfo {author} {\bibfnamefont {a.~H.}\ \bibnamefont {{Castro Neto}}},
  \bibinfo {author} {\bibfnamefont {A.}~\bibnamefont {Reina}}, \bibinfo
  {author} {\bibfnamefont {J.}~\bibnamefont {Kong}}, \ and\ \bibinfo {author}
  {\bibfnamefont {E.~Y.}\ \bibnamefont {Andrei}},\ }\href {\doibase
  10.1038/nphys1463} {\bibfield  {journal} {\bibinfo  {journal} {Nature
  Physics}\ }\textbf {\bibinfo {volume} {6}},\ \bibinfo {pages} {109} (\bibinfo
  {year} {2009})}\BibitemShut {NoStop}%
\bibitem [{\citenamefont {Luican}\ \emph {et~al.}(2011)\citenamefont {Luican},
  \citenamefont {Li}, \citenamefont {Reina}, \citenamefont {Kong},
  \citenamefont {Nair}, \citenamefont {Novoselov}, \citenamefont {Geim},\ and\
  \citenamefont {Andrei}}]{Luican2011}%
  \BibitemOpen
  \bibfield  {author} {\bibinfo {author} {\bibfnamefont {A.}~\bibnamefont
  {Luican}}, \bibinfo {author} {\bibfnamefont {G.}~\bibnamefont {Li}}, \bibinfo
  {author} {\bibfnamefont {A.}~\bibnamefont {Reina}}, \bibinfo {author}
  {\bibfnamefont {J.}~\bibnamefont {Kong}}, \bibinfo {author} {\bibfnamefont
  {R.~R.}\ \bibnamefont {Nair}}, \bibinfo {author} {\bibfnamefont {K.~S.}\
  \bibnamefont {Novoselov}}, \bibinfo {author} {\bibfnamefont {a.~K.}\
  \bibnamefont {Geim}}, \ and\ \bibinfo {author} {\bibfnamefont {E.~Y.}\
  \bibnamefont {Andrei}},\ }\href {\doibase 10.1103/PhysRevLett.106.126802}
  {\bibfield  {journal} {\bibinfo  {journal} {Physical review letters}\
  }\textbf {\bibinfo {volume} {106}},\ \bibinfo {pages} {126802} (\bibinfo
  {year} {2011})}\BibitemShut {NoStop}%
\bibitem [{\citenamefont {Wong}\ \emph {et~al.}(2015)\citenamefont {Wong},
  \citenamefont {Wang}, \citenamefont {Jung}, \citenamefont {Pezzini},
  \citenamefont {DaSilva}, \citenamefont {Tsai}, \citenamefont {Jung},
  \citenamefont {Khajeh}, \citenamefont {Kim}, \citenamefont {Lee},
  \citenamefont {Kahn}, \citenamefont {Tollabimazraehno}, \citenamefont
  {Rasool}, \citenamefont {Watanabe}, \citenamefont {Taniguchi}, \citenamefont
  {Zettl}, \citenamefont {Adam}, \citenamefont {MacDonald},\ and\ \citenamefont
  {Crommie}}]{Wong2015}%
  \BibitemOpen
  \bibfield  {author} {\bibinfo {author} {\bibfnamefont {D.}~\bibnamefont
  {Wong}}, \bibinfo {author} {\bibfnamefont {Y.}~\bibnamefont {Wang}}, \bibinfo
  {author} {\bibfnamefont {J.}~\bibnamefont {Jung}}, \bibinfo {author}
  {\bibfnamefont {S.}~\bibnamefont {Pezzini}}, \bibinfo {author} {\bibfnamefont
  {A.~M.}\ \bibnamefont {DaSilva}}, \bibinfo {author} {\bibfnamefont {H.-Z.}\
  \bibnamefont {Tsai}}, \bibinfo {author} {\bibfnamefont {H.~S.}\ \bibnamefont
  {Jung}}, \bibinfo {author} {\bibfnamefont {R.}~\bibnamefont {Khajeh}},
  \bibinfo {author} {\bibfnamefont {Y.}~\bibnamefont {Kim}}, \bibinfo {author}
  {\bibfnamefont {J.}~\bibnamefont {Lee}}, \bibinfo {author} {\bibfnamefont
  {S.}~\bibnamefont {Kahn}}, \bibinfo {author} {\bibfnamefont {S.}~\bibnamefont
  {Tollabimazraehno}}, \bibinfo {author} {\bibfnamefont {H.}~\bibnamefont
  {Rasool}}, \bibinfo {author} {\bibfnamefont {K.}~\bibnamefont {Watanabe}},
  \bibinfo {author} {\bibfnamefont {T.}~\bibnamefont {Taniguchi}}, \bibinfo
  {author} {\bibfnamefont {A.}~\bibnamefont {Zettl}}, \bibinfo {author}
  {\bibfnamefont {S.}~\bibnamefont {Adam}}, \bibinfo {author} {\bibfnamefont
  {A.~H.}\ \bibnamefont {MacDonald}}, \ and\ \bibinfo {author} {\bibfnamefont
  {M.~F.}\ \bibnamefont {Crommie}},\ }\href {\doibase
  10.1103/PhysRevB.92.155409} {\bibfield  {journal} {\bibinfo  {journal}
  {Physical Review B}\ }\textbf {\bibinfo {volume} {92}},\ \bibinfo {pages}
  {155409} (\bibinfo {year} {2015})}\BibitemShut {NoStop}%
\bibitem [{\citenamefont {Yin}\ \emph {et~al.}(2015)\citenamefont {Yin},
  \citenamefont {Qiao}, \citenamefont {Zuo}, \citenamefont {Li},\ and\
  \citenamefont {He}}]{Yin2015}%
  \BibitemOpen
  \bibfield  {author} {\bibinfo {author} {\bibfnamefont {L.-J.}\ \bibnamefont
  {Yin}}, \bibinfo {author} {\bibfnamefont {J.-B.}\ \bibnamefont {Qiao}},
  \bibinfo {author} {\bibfnamefont {W.-J.}\ \bibnamefont {Zuo}}, \bibinfo
  {author} {\bibfnamefont {W.-T.}\ \bibnamefont {Li}}, \ and\ \bibinfo {author}
  {\bibfnamefont {L.}~\bibnamefont {He}},\ }\href {\doibase
  10.1103/PhysRevB.92.081406} {\bibfield  {journal} {\bibinfo  {journal}
  {Physical Review B}\ }\textbf {\bibinfo {volume} {92}},\ \bibinfo {pages}
  {081406} (\bibinfo {year} {2015})}\BibitemShut {NoStop}%
\bibitem [{\citenamefont {Stroscio}\ \emph {et~al.}(1986)\citenamefont
  {Stroscio}, \citenamefont {Feenstra},\ and\ \citenamefont
  {Fein}}]{Stroscio1986}%
  \BibitemOpen
  \bibfield  {author} {\bibinfo {author} {\bibfnamefont {J.}~\bibnamefont
  {Stroscio}}, \bibinfo {author} {\bibfnamefont {R.}~\bibnamefont {Feenstra}},
  \ and\ \bibinfo {author} {\bibfnamefont {A.}~\bibnamefont {Fein}},\ }\href
  {\doibase 10.1103/PhysRevLett.57.2579} {\bibfield  {journal} {\bibinfo
  {journal} {Physical review letters}\ }\textbf {\bibinfo {volume} {57}},\
  \bibinfo {pages} {2579} (\bibinfo {year} {1986})}\BibitemShut {NoStop}%
\bibitem [{\citenamefont {Huang}\ \emph {et~al.}(2015)\citenamefont {Huang},
  \citenamefont {Song}, \citenamefont {Webb}, \citenamefont {Fang},
  \citenamefont {Chang}, \citenamefont {Moodera}, \citenamefont {Kaxiras},\
  and\ \citenamefont {Hoffman}}]{Huang2015}%
  \BibitemOpen
  \bibfield  {author} {\bibinfo {author} {\bibfnamefont {D.}~\bibnamefont
  {Huang}}, \bibinfo {author} {\bibfnamefont {C.-L.}\ \bibnamefont {Song}},
  \bibinfo {author} {\bibfnamefont {T.~a.}\ \bibnamefont {Webb}}, \bibinfo
  {author} {\bibfnamefont {S.}~\bibnamefont {Fang}}, \bibinfo {author}
  {\bibfnamefont {C.-Z.}\ \bibnamefont {Chang}}, \bibinfo {author}
  {\bibfnamefont {J.~S.}\ \bibnamefont {Moodera}}, \bibinfo {author}
  {\bibfnamefont {E.}~\bibnamefont {Kaxiras}}, \ and\ \bibinfo {author}
  {\bibfnamefont {J.~E.}\ \bibnamefont {Hoffman}},\ }\href {\doibase
  10.1103/PhysRevLett.115.017002} {\bibfield  {journal} {\bibinfo  {journal}
  {Physical review letters}\ }\textbf {\bibinfo {volume} {115}},\ \bibinfo
  {pages} {017002} (\bibinfo {year} {2015})}\BibitemShut {NoStop}%
\bibitem [{\citenamefont {Brihuega}\ \emph {et~al.}(2012)\citenamefont
  {Brihuega}, \citenamefont {Mallet}, \citenamefont {Gonz\'{a}lez-Herrero},
  \citenamefont {{Trambly de Laissardi\`{e}re}}, \citenamefont {Ugeda},
  \citenamefont {Magaud}, \citenamefont {G\'{o}mez-Rodr\'{\i}guez},
  \citenamefont {Yndur\'{a}in},\ and\ \citenamefont {Veuillen}}]{Brihuega2012}%
  \BibitemOpen
  \bibfield  {author} {\bibinfo {author} {\bibfnamefont {I.}~\bibnamefont
  {Brihuega}}, \bibinfo {author} {\bibfnamefont {P.}~\bibnamefont {Mallet}},
  \bibinfo {author} {\bibfnamefont {H.}~\bibnamefont {Gonz\'{a}lez-Herrero}},
  \bibinfo {author} {\bibfnamefont {G.}~\bibnamefont {{Trambly de
  Laissardi\`{e}re}}}, \bibinfo {author} {\bibfnamefont {M.~M.}\ \bibnamefont
  {Ugeda}}, \bibinfo {author} {\bibfnamefont {L.}~\bibnamefont {Magaud}},
  \bibinfo {author} {\bibfnamefont {J.~M.}\ \bibnamefont
  {G\'{o}mez-Rodr\'{\i}guez}}, \bibinfo {author} {\bibfnamefont
  {F.}~\bibnamefont {Yndur\'{a}in}}, \ and\ \bibinfo {author} {\bibfnamefont
  {J.-Y.}\ \bibnamefont {Veuillen}},\ }\href {\doibase
  10.1103/PhysRevLett.109.196802} {\bibfield  {journal} {\bibinfo  {journal}
  {Physical review letters}\ }\textbf {\bibinfo {volume} {109}},\ \bibinfo
  {pages} {196802} (\bibinfo {year} {2012})}\BibitemShut {NoStop}%
\bibitem [{\citenamefont {Streda}(1982)}]{Streda1982}%
  \BibitemOpen
  \bibfield  {author} {\bibinfo {author} {\bibfnamefont {P.}~\bibnamefont
  {Streda}},\ }\href
  {http://iopscience.iop.org/article/10.1088/0022-3719/15/36/006/meta}
  {\bibfield  {journal} {\bibinfo  {journal} {Journal of Physics C: Solid State
  Physics}\ }\textbf {\bibinfo {volume} {15}},\ \bibinfo {pages} {1299}
  (\bibinfo {year} {1982})}\BibitemShut {NoStop}%
\bibitem [{\citenamefont {Wei\ss~e}\ \emph {et~al.}(2006)\citenamefont
  {Wei\ss~e}, \citenamefont {Wellein}, \citenamefont {Alvermann},\ and\
  \citenamefont {Fehske}}]{Weiße2006a}%
  \BibitemOpen
  \bibfield  {author} {\bibinfo {author} {\bibfnamefont {A.}~\bibnamefont
  {Wei\ss~e}}, \bibinfo {author} {\bibfnamefont {G.}~\bibnamefont {Wellein}},
  \bibinfo {author} {\bibfnamefont {A.}~\bibnamefont {Alvermann}}, \ and\
  \bibinfo {author} {\bibfnamefont {H.}~\bibnamefont {Fehske}},\ }\href
  {\doibase 10.1103/RevModPhys.78.275} {\bibfield  {journal} {\bibinfo
  {journal} {Reviews of Modern Physics}\ }\textbf {\bibinfo {volume} {78}},\
  \bibinfo {pages} {275} (\bibinfo {year} {2006})}\BibitemShut {NoStop}%
\bibitem [{\citenamefont {{Di Napoli}}\ \emph {et~al.}(2016)\citenamefont {{Di
  Napoli}}, \citenamefont {Polizzi},\ and\ \citenamefont {Saad}}]{Napoli2016}%
  \BibitemOpen
  \bibfield  {author} {\bibinfo {author} {\bibfnamefont {E.}~\bibnamefont {{Di
  Napoli}}}, \bibinfo {author} {\bibfnamefont {E.}~\bibnamefont {Polizzi}}, \
  and\ \bibinfo {author} {\bibfnamefont {Y.}~\bibnamefont {Saad}},\ }\href
  {\doibase 10.1002/nla.2048} {\bibfield  {journal} {\bibinfo  {journal}
  {Numerical Linear Algebra with Applications}\ }\textbf {\bibinfo {volume}
  {23}},\ \bibinfo {pages} {674} (\bibinfo {year} {2016})}\BibitemShut
  {NoStop}%
\end{thebibliography}%
	
\end{document}